\documentclass[10pt]{svjour3}
\usepackage{fullpage}
\usepackage{times}
\usepackage{graphicx}
\usepackage{epsfig}
\usepackage{amsmath}
\usepackage{amssymb}

\twocolumn

\begin{document}

\newcommand{\ignore}[1]{}
\newcommand{\db}[1]{\textbf{[[[db:#1]]]}}
\newcommand{\dd}[1]{\textbf{[[[dd:#1]]]}}

\newcommand{\bp}[1]{\textbf{[[[bp:#1]]]}}
\newcommand{\tr}[1]{\textbf{[[[\bf tr:#1]]]}}

\newcommand{\namedref}[2]{{#1~\ref{#2}}}
\newcommand{\sectionref}[1]{\namedref{Section}{#1}}
\newcommand{\appendixref}[1]{\namedref{Appendix}{#1}}
\newcommand{\subsectionref}[1]{\namedref{Subsection}{#1}}
\newcommand{\theoremref}[1]{\namedref{Theorem}{#1}}
\newcommand{\defref}[1]{\namedref{Definition}{#1}}
\newcommand{\figureref}[1]{\namedref{Figure}{#1}}
\newcommand{\figref}[1]{\namedref{Figure}{#1}}
\newcommand{\claimref}[1]{\namedref{Claim}{#1}}
\newcommand{\lemmaref}[1]{\namedref{Lemma}{#1}}
\newcommand{\tableref}[1]{\namedref{Table}{#1}}
\newcommand{\corollaryref}[1]{\namedref{Corollary}{#1}}
\newcommand{\propertyref}[1]{\namedref{Property}{#1}}
\newcommand{\appref}[1]{\namedref{Appendix}{#1}}
\newcommand{\propref}[1]{\namedref{Proposition}{#1}}
\newcommand{\algref}[1]{\namedref{Algorithm}{#1}}

\newcommand{\eqnref}[1]{{Eq.~\eqref{#1}}}

\newcommand{\comment}[1]  {}
\newcommand\ie{{\textsl{i.e.\,}}}
\newcommand\eg{{\textsl{e.g.\,}}}
\newcommand\etal{{\textsl{et al.\,}}}
\def\BE{\begin{equation}}
\def\EE{\end{equation}}
\def\BEA{\begin{eqnarray}}
\def\EEA{\end{eqnarray}}
\newcommand{\cut}[1]{{}}
\newcommand\va{{\bf a}} 
\newcommand\vb{{\bf b}}
\newcommand\vc{{\bf c}}
\newcommand\vd{{\bf d}}
\newcommand\ve{{\bf e}}
\newcommand\vf{{\bf f}}
\newcommand\vg{{\bf g}}
\newcommand\vh{{\bf h}}
\newcommand\vi{{\bf i}}
\newcommand\vj{{\bf j}}
\newcommand\vk{{\bf k}}
\newcommand\vl{{\bf l}}
\newcommand\vm{{\bf m}}
\newcommand\vn{{\bf n}}
\newcommand\vo{{\bf o}}
\newcommand\vp{{\bf p}}
\newcommand\vq{{\bf q}}
\newcommand\vr{{\bf r}}
\newcommand\vs{{\bf s}}
\newcommand\vt{{\bf t}}
\newcommand\vu{{\bf u}}
\newcommand\vv{{\bf v}}
\newcommand\vw{{\bf w}}
\newcommand\vx{{\bf x}}
\newcommand\vy{{\bf y}}
\newcommand\vz{{\bf z}}
\newcommand\mA{{\bf A}} 
\newcommand\mB{{\bf B}}
\newcommand\mC{{\bf C}}
\newcommand\mD{{\bf D}}
\newcommand\mE{{\bf E}}
\newcommand\mF{{\bf F}}
\newcommand\mG{{\bf G}}
\newcommand\mH{{\bf H}}
\newcommand\mI{{\bf I}}
\newcommand\mJ{{\bf J}}
\newcommand\mK{{\bf K}}
\newcommand\mL{{\bf L}}
\newcommand\mM{{\bf M}}
\newcommand\mN{{\bf N}}
\newcommand\mO{{\bf O}}
\newcommand\mP{{\bf P}}
\newcommand\mQ{{\bf Q}}
\newcommand\mR{{\bf R}}
\newcommand\mS{{\bf S}}
\newcommand\mT{{\bf T}}
\newcommand\mU{{\bf U}}
\newcommand\mV{{\bf V}}
\newcommand\mW{{\bf W}}
\newcommand\mX{{\bf X}}
\newcommand\mY{{\bf Y}}
\newcommand\mZ{{\bf Z}}

\title{Peer-to-Peer Secure Multi-Party Numerical Computation\\
Facing Malicious Adversaries\thanks{Earlier version of the
material in this paper was presented in part at the
        8th IEEE Peer-to-Peer Computing Conference, Sept. 2008, Aachen, Germany.}
        \thanks{Danny Dolev and Tzachy Reinman where supported by The Israel Science Foundation
  (grant No.~1685/07). Benny Pinkas was supported by The Israel Science Foundation
  (grant No.~860/06).}}
\author{Danny Bickson, Tzachy Reinman, Danny Dolev and Benny Pinkas}
\institute{
Danny Bickson\\
IBM Haifa Research Lab,\\Mount Carmel, Haifa 31905, Israel.\\ dannybi@il.ibm.com\\
\and
Tzachy Reinman and Danny Dolev\\
School of Computer Science and Engineering\\
The Hebrew University of Jerusalem,
Jerusalem 91904, Israel.\\
\{reinman,dolev\}@cs.huji.ac.il\\
\and Benny Pinkas\\
Dept. of Computer Science,\\ University of Haifa, Mount Carmel, Haifa 31905, Israel.\\
benny@pinkas.net\\
}

\maketitle
\thispagestyle{empty}

\begin{abstract}
We propose an efficient framework for enabling secure multi-party
numerical computations in a Peer-to-Peer network. This problem
arises in a range of applications such as collaborative filtering,
distributed computation of trust and reputation, monitoring and
other tasks, where the computing nodes is expected to
preserve  the privacy of their inputs while performing  a joint
computation of a certain function.

Although there is a rich literature in the field of distributed
systems security concerning secure multi-party computation, in
practice it is hard to deploy those methods in very large scale
Peer-to-Peer networks. In this work, we try to bridge the
gap between theoretical algorithms in the security domain, and a practical Peer-to-Peer deployment.

We consider two security models. The first is the semi-honest model where peers correctly follow the protocol, but
try to reveal private information. We provide three possible schemes for secure multi-party numerical
computation for this model and identify a single light-weight scheme which outperforms the others. Using extensive simulation results over real Internet topologies, we demonstrate that our scheme is scalable to very large networks, with up to millions of nodes.

The second model we consider is the malicious peers model, where peers can behave arbitrarily,
deliberately trying to affect the results of the computation as well as compromising the privacy of other peers.
For this model we provide a fourth scheme to defend the execution of the computation against the malicious peers. The proposed scheme has a higher complexity relative to the semi-honest model. Overall, we provide the Peer-to-Peer
network designer a set of tools to choose from, based on the desired level of security.

\end{abstract}

\section{Introduction}
We consider the problem of performing a joint numerical
computation of some function over a Peer-to-Peer network. Such
problems arise in many applications, for example, when computing
distributively trust~\cite{EigenTrust}, ranking of nodes and data
items~\cite{p2p-rating}, clustering~\cite{EWSN08}, collaborative
filtering~\cite{KorenCF,PP2}, factor analysis~\cite{Canny} etc.
The aim of {\em secure multi-party computation} is to enable
parties to carry out such distributed computing tasks in a secure
manner. Whereas distributed computing classically deals with
questions of computing under the threat of machine crashes and
other inadvertent faults, secure multi-party computation is
concerned with the possibility of deliberate malicious behavior by
some adversarial entity. That is, it is assumed that a protocol
execution may come under attack by an external entity, or even by
a subset of the participating parties. The aim of this attack may
be to learn private information or cause the result of the
computation to be incorrect. Thus, two central requirements on any
secure computation protocol are privacy and correctness. The
privacy requirement states that nothing should be learned beyond
what is absolutely necessary; more exactly, parties should learn
their designated output and nothing else. The correctness
requirement states that each party should receive its correct
output. Therefore, the adversary must not be able to cause the
result of the computation to deviate from the function that the
parties had set out to compute.

In this paper, we consider only functions which are built using the
algebraic primitives of addition, substraction and multiplication. In
particular, we focus on numerical methods which are computed
distributively in a Peer-to-Peer network, where in each iteration,
every node interacts with a subset of its neighbors by sending scalar
messages, and computing a weighted sum of the messages that it
receives. Examples of such functions are belief
propagation~\cite{BibDB:BookPearl}, EM (expectation
maximization)~\cite{Canny}, Power method~\cite{EigenTrust}, separable
functions~\cite{Separable}, gradient descent methods~\cite{PP3} and
linear iterative algorithms for solving systems of linear
equations~\cite{BibDB:BookBertsekasTsitsiklis}.  As a specific
example, we describe the Jacobi algorithm for computing such functions
in detail in \sectionref{jacobi}.

There is a rich body of research on secure computation, starting
with the seminal work of Yao~\cite{Yao}. Part of this research is
concerned with the design of {\em generic} secure protocols that
can be used for computing any function (for example, Yao's
work~\cite{Yao} for the case of two participants, and
e.g.~\cite{BGW,GMW} for solutions for the case of multiple
participants).  There are several works concerning the {\em
implementation} of generic protocols for secure computation. For
example, FairPlay~\cite{Fairplay} is a system for secure two-party
computation, and FairPlayMP~\cite{FairPlayMP} is a different
system for secure computation by more than two parties.  These two
systems are based (like Yao's protocol) on reducing any function
to a representation as a Boolean circuit and computing the
resulting Boolean circuit securely. Our approach is much more
efficient, at the cost of supporting only a subset of the
functions the FairPlay system can compute.

A different line of work studies secure protocols for computing
specific functions (rather than generic protocols for computing any
function). Of particular interest for us are works that add
a privacy preserving layer to the computation of functions such as the
factor analysis learning problem (for which~\cite{Canny} describes a
secure multi-party protocol using homomorphic encryption),
computing trust in a Peer-to-Peer network (for which~\cite{EigenTrust}
suggests a solution using a trusted third party), or the work
of~\cite{PP3}, which is closely related to our work, but is limited to
two parties.

Most  previous solutions for secure multi-party computation suffer
from one of the following drawbacks: (1) they provide a
centralized solution where all information is shipped to a single
computing node, and/or (2) require communication between all
participants in the protocol, and/or (3) require the use of asymmetric
encryption, which is costly. In this work, we investigate
 secure computation in a Peer-to-Peer setting, where each
node is only connected to some of the other nodes (its neighbors).
We examine different possible distributed approaches, and out of the them
we identify a single approach, which is theoretically
secure and at the same time efficient and scalable.

Security is often based on the assumption that there is an upper
bound on the {\em global} number of malicious participants. In our
setting, we consider the number of malicious nodes in each {\em
local vicinity}. Furthermore, most of the existing algorithms
scale to tens or hundreds of nodes, at the most. In this work, we address
the problem in a setting of a large Peer-to-Peer network, with
millions of nodes and hundreds of millions of communication links.
Unlike most of the previous work, we have performed a {\em very
large scale simulation}, using real Internet topologies, demonstrating that
our approach is applicable to real network settings.

As an example for applications of our framework, we take the
neighborhood based collaborative filtering~\cite{KorenCF}. This
algorithm is a recent state-of-the-art algorithm. There are two
challenges in adapting this algorithm to a Peer-to-Peer network.
First, the algorithm is centralized and we propose a method to
distribute it. Second, we add a privacy preserving layer, so no
information about personal ranking is revealed during the process
of computation.

The paper is organized as follows. In \sectionref{model} we
formulate our problem model. In \sectionref{crypto} we give a
brief background of cryptographic primitives that are used in our
schemes. \sectionref{const} outlines our novel construction for
the semi-honest model. In \sectionref{crypto2} we review
cryptographic primitives needed for extending our construction to
support the malicious adversary model.
\sectionref{malicious} presents our extended construction for the
malicious adversary model. Example collaborative filtering
application is given in \sectionref{jacobi}. Large scale
simulations are presented in \sectionref{exp}. We conclude in
\sectionref{Conclusion}.

We use the following notations: $T$ stands for a vector or matrix
transpose, the symbols $\{\cdot\}_{i}$ and $\{\cdot\}_{ij}$ denote
entries of a vector and matrix, respectively. $N_i$ is the set of neighboring nodes to node $i$. The spectral radius
$\rho(\mB)\triangleq\max_{1\leq i\leq s}(|\lambda_{i}|)$, where
$\lambda_{1},\ldots, \lambda_{s}$ are the eigenvalues of a matrix
$\mB$.

\section{Our Model}
\label{model} Given a Peer-to-Peer network graph $G=(V, E)$ with
$|V| = n$ nodes and $|E| = e$ edges, we would like to perform a
joint iterative computation. Each node $i$ starts with a
scalar\footnote{An extension to the vector case is immediate, we
omit it for the clarify of description.} state $x_i^0 \in
\mathbb{R}$, and on each round sends messages to a subset of its
neighbors. We denote a message sent from node $i$ to node $j$ at
round $r$ as $m_{ij}^r$.

Let $N_i$ denote the set of neighboring nodes of $i$.  Denote the
neighbors of node $i$ as $n_{i_1},n_{i_2},\ldots,n_{i_k}$, where
$k=|N_i|$.  We assume, wlog, that each node sends a message to a
subset of its neighbors (possibly including itself.)  On each round $r = 1,2,\cdots$, node $i$
computes, based on the messages it received, a function $f:
R^{k+1} \rightarrow R^{k+1}$,
\small
\begin{gather*}\begin{split}\langle m_{i~i}^r, m^{r}_{i~n_{i_1}}, \cdots, m^{r}_{i~n_{i_k}}\rangle &=
f(m_{i~i}^{r-1}, m^{r-1}_{n_{i_1} i}, \cdots , m^{r-1}_{n_{i_k}
i})\;.
\end{split}\end{gather*}
\normalsize

Namely, the function gets as input the initial state (which is
denoted as a self message $m_{i~i}$) and all the received neighbor
messages of this round, and outputs a new state and messages to be
sent to a subset of the neighbors at the next round. The iterative
algorithms run either a predetermined number of rounds, or until
convergence is detected locally. Whenever the reference to the
round number is clear from the text, the round numbers are omitted
to simplify our notations.

In this paper, we are only interested in functions $f$ that
compute weighted sums on each iteration. Next we show that there
is a variety of such numerical methods. Our goal is to add a
privacy preserving layer to the distributed computation, such that
the only information learned by a node is its share of the output.

In \sectionref{const} we use the semi-honest adversaries model: in this model (common in
cryptographic research of secure computation) even corrupted parties
are assumed to correctly follow the protocol specification. However,
the adversary obtains the internal states of all the corrupted parties
(including the transcript of all the messages received), and attempts
to use this information to learn information that should remain
private.

Security against semi-honest adversaries might be
 justified if the parties participating in the protocol are somewhat
 trusted, or if we trust the participating parties at the time they
 execute the protocol, but suspect that at a later time an adversary
 might corrupt them and get hold of the transcript of the information
 received in the protocol.

\sectionref{malicious} extends our construction to the ``malicious adversary'',
which can behave arbitrarily.  We note that protocols secure against malicious adversaries are
 considerably more costly than their semi-honest counterparts. For
 example, the generic method of obtaining security against malicious
  adversaries is through the GMW compiler~\cite{GMW} which adds a
  zero-knowledge proof for every step of the protocol.


We define a configurable local system parameter $d_i$, where $d_i-1$ is  the
maximum number of nodes in the local vicinity of node $i$
(direct neighbors of node $i$) that might be corrupted. Whenever this
assertion is violated, the security of our proposed scheme is
affected. This is a stronger requirement from our system, relative
to the traditional global bound on the number of adversarial
nodes.


\section{Cryptographic primitives for the semi-host model}
\label{crypto} We compare several existing approaches from the
literature of secure multi-party computation and discuss their
relevance to Peer-to-Peer networks.

\subsection{Random perturbations}
The random additive perturbation method attempts to preserve the
privacy of the data by modifying values of the sensitive
attributes using a randomized process
(see~\cite{AS,DiNi,RandomNoise}). In this approach, the node sends
a value $u_i + v$, where $u_i$ is the original scalar message, and
$v$ is a random value drawn from a certain distribution $V$. In
order to perturb the data, $n$ independent samples $v_1, v_2,
\cdots , v_n$, are drawn from a distribution $V$. The owners of
the data provide the perturbed values $u_1+v_1, u_2+v_2, \cdots ,
u_n+v_n$ and the cumulative distribution function $FV(r)$ of $V$.
The goal is to use these values, instead of the original ones, in
the computation. (It is easy to see, for example, that if the
expected value of $V$ is $0$, then the expectation of the sum of
the $u_i+v_i$ values is equal to the expectation of the $u_i$
values.) The hope is that by adding random noise to the individual
data points it is possible to hide the individual values.

\ignore{ The $n$ original data values $u_1, u_2, \cdots , u_n$ are
  viewed as realizations of $n$ independent and identically
  distributed (i.i.d.) random variables $U_i, i = 1, 2, \cdots , n$,
  each with the same distribution as that of a random variable U.  In
  order to perturb the data, $n$ independent samples $v_1, v_2, \cdots
  , v_n$, are drawn from a distribution $V$ . The owner of the data
  provides the perturbed values $u_1+v_1, u_2+v_2, \cdots , u_n+v_n$
  and the cumulative distribution function $FV(r)$ of $V$. The
  reconstruction problem is to estimate the distribution $FU(x)$ of
  the original data, from the perturbed data.
}

The random perturbation model is limited. It supports only
addition operations, and it was shown in~\cite{DiNi} that this
approach can ensure very limited privacy guarantees. We only
present
this method
for
comparing its running time with the other protocols.

\subsection{Shamir's Secret Sharing (SSS)}
Secret sharing is a fundamental primitive of cryptographic
protocols. We will describe the  secret sharing scheme of
Shamir~\cite{SSS}. The scheme works over a field $F$, and it is
assumed that the secret $s$ is an element in that field. In a
$k$-out-of-$n$ secret sharing the owner of a secret wishes to
distribute it among $n$ players such that any subset of at least $k$ of
them is able to recover the secret, while no subset of up to $k-1$
players is able to learn any information about the secret. (In the
application described in this paper each player will be a node in the
network.)

In order to distribute the secret, its owner chooses a random
polynomial $P()$ of degree $k-1$, subject to the constraint that
$P(0)=s$. This is done by choosing random coefficients
$a_1,\ldots,a_{k-1}$ and defining the polynomial as
$P(x)=s+\sum_{i=1}^{k-1}a_ix^i$. Each player is associated with an
identity in the field (denoted $x_1,\ldots,x_n$\/ for players
$1,\ldots,n$, respectively). The share that player $i$ receives is the
value $P(x_i)$, namely the value of the polynomial evaluated at the
point $x_i$.  It is easy to see that any $k$ players can recover the
secret, since they have $k$ values of the polynomial and can therefore
interpolate it and compute its free coefficient $s$. It is also not
hard to see that any set of up to $k-1$ players does not learn any
information about $s$, since any value of $s$ has a probability of
$1/|F|$ of resulting in a polynomial which agrees with the values that
the players have.
\subsection{Homomorphic encryption}
A homomorphic encryption scheme is an encryption scheme that
allows certain algebraic operations to be carried out on the
encrypted plaintext, by applying an efficient operation to the
corresponding ciphertext (without knowing the decryption key!). In
particular, we will be interested in additively homomorphic
encryption schemes: Here, the message space is a ring (or a
field). There exists an efficient algorithm $+_{pk}$ whose input
is the public key of the encryption scheme and two ciphertexts,
and whose output is $E_{pk}(m_1) +_{pk} E_{pk}(m_2) = E_{pk}(m_1 +
m_2)$. (Namely, this algorithm computes, given the public key and
two ciphertexts, the encryption of the sum of the plaintexts of
two ciphertexts.) There is also an efficient algorithm
$\cdot_{pk}$, whose input consists of the public key of the
encryption scheme, a ciphertext, and a constant $c$ in the ring,
and whose output is  $c\cdot_{pk} E_{pk}(m) = E_{pk}(c \cdot m)$.

We will also  require that the encryption scheme has semantic
security. An efficient implementation of an additive homomorphic
encryption scheme with semantic security was given by
Paillier~\cite{Paillier}. In this cryptosystem the encryption of a
plaintext from $[1;N]$, where N is a RSA modulus, requires two
exponentiations modulo $N^2$. Decryption requires a single
exponentiation. We will use this encryption scheme in our work.

\subsubsection{Paillier encryption}
We describe in a nutshell the Paillier cryptosystem. Fuller
details are found on~\cite{Paillier}.
\begin{itemize}
    \item {\bf Key generation} Generate two large primes $p$
and $q$. The secret key $sk$ is $\lambda = lcm(p - 1, q - 1)$. The
public key $pk$ includes $N = pq$ and $g  \in \mathbb{Z}_{N^2}$
such that $g \equiv \mbox{  }1 \mbox{  } \mod N$.
    \item {\bf Encryption} Encrypt a message $m \in \mathbb{Z}_{N}$ with randomness
$r \in \mathbb{Z}_{N^2}^*$ and public key $pk$ as $c = g^mr^N \mod
N^2$.
    \item {\bf Decryption} Decrypt a ciphertext $c \in \mathbb{Z}_{N^2}^*$.
    Decryption is done using: $\frac{L(c^\lambda \mod N^2)}{L(g^\lambda \mod
    N^2)} \mod N$ where $L(x) = (x - 1)/N$.
\end{itemize}


\section{Our construction}
\label{sec:semi_honest} \label{const} The main observation we make
is that numerous distributed numerical methods compute in each
node a weighted sum of scalars $m_{ji}$, received from neighboring
nodes, namely \BE \label{ws} \sum_{j\in N_i} a_{ji} m_{ji}\;, \EE
where the weight coefficients $a_{ji}$  are known constants. This
simple building block captures the behavior of multiple numerical
methods. By showing ways to compute this weighted sum securely,
our framework can support many of those numerical methods. In this
section we introduce three possible approaches for performing the
weighted sum computation.

In \sectionref{Jacobi} we give an example of the Jacobi algorithm
which computes such a weighted sum on each iteration.

\subsection{A Construction Based on Random Perturbations}
\ignore{ In this method, we only support distributed calculations
of sums. } In each iteration of the algorithm, whenever a node $j$
needs to send a value $m_{ji}$ to a neighboring node $i$, the node
$j$ generates a random number $r_{ji}$ using the GMP
library~\cite{GMP}, from a probability distribution with zero
mean. It then sends the value $m_{ji}+r_{ji}$ to the other node
$i$.
 As the number of neighbors increases,
the computed noisy sum $\sum_{j\in N_i} (m_{ji}+r_{ji})$ converges
to the actual sum $\sum_{j\in N_i} m_{ji}$.

When the node $i$ computes a weighted sum of the messages it
received as in equation~\ref{ws}, it multiplies each incoming
message by the corresponding weight. The computed noisy sum
$\sum_{j\in N_i} a_{ji}(m_{ji}+r_{ji})$
 converges to the
actual sum  $\sum_{j\in N_i} a_{ji}m_{ji}$.

We note again that this method is considered mainly for a comparison
of its  running
time with those of  the other methods.

\subsection{A Construction Based on Homomorphic Encryption}
\label{Homomorphic}
%

We chose to utilize the Paillier encryption scheme, which is an
efficient realization of an additive homomorphic encryption scheme
with
semantic security.\\
\\
{\bf Key generation: } We use the threshold version of the
Paillier encryption scheme described in~\cite{Paillier2}. In this
scheme, a trusted third party generates for each node $i$ private
and public key pairs.\footnote{It is also possible to generate the key
  in a distributed way, without using any trusted  party. This option
  is less efficient. We  show that eventhough the usage of a
  centralized key generation process is not efficient enough, and
  therefore we have not implemented the distributed version of this protocol.}
  The public key is disseminated to all of
node $i$ neighbors. The private key $\lambda_i=prvk(i)$ is kept
secret from all nodes (including node $i$). Instead, it is split,
using secret sharing, to the neighbors of node $i$. There is a
threshold $d_i$, which is at most equal to $|N_i|$, the number of
neighbors of node $i$. The scheme ensures that any subset of $d_i$
of the neighbors of node $i$ can help it decrypt messages (without
the neighbors learning the decrypted message, or node $i$ learning
the private key). If $d_i=|N_i|$ then the private key is shared by
giving each neighbor $j$ a random value $s_{ji}$ subject to the
constraint $\sum_{j\in N_i}s_{ji} = \lambda_i=prvk(i)$. Otherwise,
if $d_i<|N_i|$ the values $s_{ji}$ are shares of a Shamir secret
sharing of $\lambda_i$. Note that fewer than $d_i$ neighbors cannot
recover the key.

Using this method, all neighboring nodes of node $i$ can send
encrypted messages using $pubk(i)$ to node $i$, while node $i$ cannot
decrypt any of these messages. It can, however,
aggregate the messages using the homomorphic property and ask a  coalition
of $d_i$ or more neighbors to help it in decrypting the sum.\\
\\
{\bf The initialization step} of this protocol is as follows:
\begin{itemize}
\item [{[H0]}] The third party creates for node $i$ a public and private key
  pair, $[pubk(i), prvk(i)]$. It sends the public key $pubk(i)$ to all
  of node $i$'s neighbors, and splits  the private key into
  shares, such that each node $i$ neighbors gets a share
  $s_{ji}$. If $d_i=|N_i|$ then
 $prvk(i)=\lambda_i = \sum_{j \in N_i} s_{ji}$. Otherwise the $s_{ji}$
 values are Shamir shares of the private key.
\end{itemize}

\noindent {\bf One round of computation: } In each round of the algorithm,
when a node $j$ would like to send a scalar value $m_{ji}$ to node
$i$ it does the following:
\begin{itemize}
    \item [{[H1]}] Encrypt the message $m_{ji}$, using node $i$ public
    key to get $C_{ji} = E_{pubk(i)}(m_{ji})$.
    \item [{[H2]}] Send the result $C_{ji}$ to node $i$.
    \item [{[H3]}] Node $i$ aggregates all the incoming message $C_{ji}$, using the homomorphic
    property to get \\$E_{pubk(i)}(\sum a_{ji}m_{ji}).$
\end{itemize}

\noindent {\bf After receiving all messages:} Node $i$'s neighbors
assist it in decrypting the result $x_i$, without revealing the
private key $prvk(i)$. This is done as follows (for the case
$d_i=|N_i|$): Recall that in a Paillier decryption node $i$ needs to
raise the result computed in [H3] to the power of its private key
$\lambda_i$.
\begin{itemize}
    \item [{[H4]}] Node $i$ sends all its neighbors the result
    computed in [H3]: $C_i = E_{pubk(i)}(\sum a_{ji}m_{ji})$.
    \item [{[H5]}] Each neighbor, computes a part of the
    decryption $w_{ji} = C_i ^{s_{ji}}$ where $s_{ji}$ are node $i$ private
    key shares computed in step [H0], and sends the result
    $w_{ji}$ to node $i$.
    \item [{[H6]}] Node $i$ multiplies all the received values to get:
    \BE \Pi_{j \in N_i} w_{ji} = C_i ^{\sum_{j \in N_i} s_{ji}} = C_i
    ^{\lambda_i} = \nonumber \EE \BE = \sum a_{ji} m_{ji} \;\; \mathrm{mod} \;\; N.
\nonumber \EE
\end{itemize}
If $d_i<|N_i|$ then the reconstruction is done using Lagrange
interpolation in the exponent, where node $i$ needs to raise each
$w_{ji}$ value by the corresponding Lagrange coefficient, and then
multiply the results.

Regarding message overhead, first we need to generate and
disseminate public and private keys. This operation requires $2e$
messages, where $e = |E|$ is the number of graph edges. In each
iteration we send the same number of messages as in the original
numerical algorithm. However, assuming a security of $\ell$ bits,
and a working precision of $d$ bits, we increase the size of the
message by a factor of $\frac{ \ell}{d}$. Finally, we add $e$
messages for obtaining the private keys parts in step [H4].

Regarding computation overhead, for each message sent, we need to
perform one Paillier encryption in step [H1]. In step [H3] the
destination node performs additional $k-1$ multiplications, and
one decryption in step [H4]. At the key generation phase, we add
generation of $n$ random polynomial and their evaluation. In step
[H4] we compute an extrapolation of those $n$ polynomials.
The security of the Paillier encryption is investigated
in~\cite{Paillier,Paillier2}, where it was shown that the system
provides semantic security.

\subsection{A Construction Based on Shamir Secret Sharing}
\label{our_SSS}
We propose a  construction based on Shamir's secret sharing,
which avoids the computation cost of asymmetric encryption. In a
nutshell, we use the neighborhood of a node for adding a privacy
preserving mechanism, where only a coalition of $d_i$ or more nodes
can reveal the content of messages sent to that node.

In each round of the algorithm, when a node $j$ would like to send
a scalar value $m_{ji}$ to node $i$ it does the following:
\begin{itemize}
    \item [{[S1]}]  Generate a random polynomial $P_{ji}$ of degree
   ~$d_i-1$, of the type $P_{ji}(x) = m_{ji} + \sum_{l=1}^{d_i-1} a_l
    x^l$\/ .
    \item [{[S2]}] For each neighbor $l$ of node $i$, create a share $C_{jil}$ of
    the polynomial $P_{ji}(x)$ by evaluating it on a single point
    $x_{l}$, namely $C_{jil} \triangleq P_{ji}(x_l)$.
    \item [{[S3]}] Send $C_{jil}$ to neighbor node $l$ of node $i$.
    \item [{[S4]}] Each neighbor $l$ of node $i$ aggregates the
      shares it received from all neighbors of node $i$  and computes the value
 $S_{li} = \sum_{j \in N_i} a_{ji} C_{jil}$. (Note
 that the result of this computation is equal to the value of a polynomial of degree $d_i-1$,
 whose free coefficient is equal to the {\em weighted} sum of all messages sent to
 node $i$ by its neighbors.)
    \item [{[S5]}] Each neighbor $l$ sends the sum $S_{li}$ to node $i$.

    \item [{[S6]}]
    Node $i$ treats the value received from node $l$ as a
      value of a polynomial of degree $d_i-1$ evaluated at the point
      $x_i$. Node $i$ interpolates $P_i(x)$ for extracting the
    free coefficient, which in this case is the weighted sum of all
    messages $\sum_{j \in N_i} a_{ji} m_{ji}$.
\end{itemize}

\ignore{ Regarding the polynomial evaluation and interpolation in
steps [S2,S7] we chose the efficient implementation of
FFTEasy~\cite{FFTEasy} which uses an iterative FFT algorithm. In
this case, the polynomial is evaluated at the points \[ x_{il} =
\exp(-\frac{2 \pi j\footnote{We note $j$ as the root of the
unity}}{k} il). \] The running time of this algorithm is $d \log
d$ for both evaluation and interpolation.}

Note that the message $m_{ji}$ sent by node $j$ remains hidden
if less than  $d_i$ neighbors of $i$ collude to learn
it (this is ensured since these neighbors learn strictly less than $d_i$ values
of a polynomial of degree $d_i-1$).
The protocol requires each node $j$ to send messages to
all other
neighbors of each of its neighbors. 
We discuss the applicability of this requirement in
\sectionref{Conclusion}.

\ignore{
\paragraph{Weighted sum} Assume that node $i$ needs to compute
the weighted sum  $\sum_{j\in N_i} a_{ij} m_{ji}$, where the
$a_{ij}$ values are constants which are known to nodes $i$ and
$j$. Then in step S4 of the protocol, node $l$ computes the sum
 $S_{li} = \sum_{j \in N_i} a_j P_{ji}(x_{il})$. The rest of the
 protocol remains as before.
}

\ignore{
Regarding the number of messages sent, we have the same overhead
as in the homomorphic encryption scheme. Note, that the size of
the messages is not increased.
}

\begin{figure}
  \includegraphics[width=200pt]{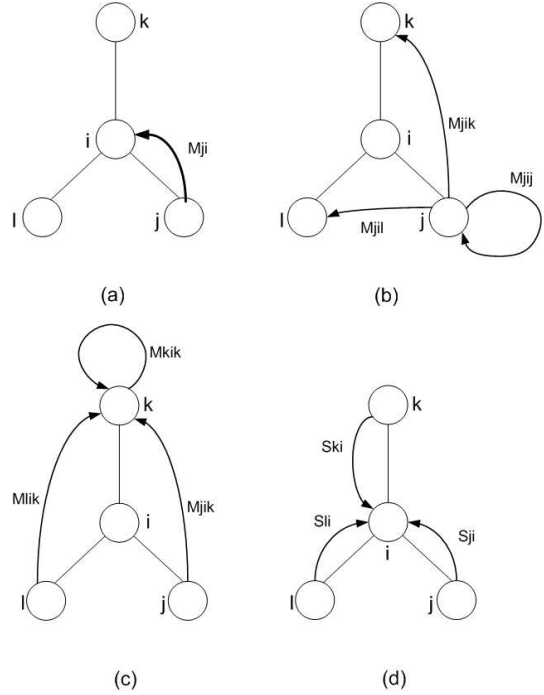}\\
  \caption{Schematic message flow in the proposed methods. The task of node $i$ is to compute the sum of all messages: $m_{ki} + m_{ji} + m_{li}$ (a) describes a message sent from $j$ to $i$ using random perturbation.
  (b) describes steps [S3] in our SSS scheme, where the same message $m_{ji}$ is split into shares sent to all of $i$ neighbors.
  (c) describes steps [S4] in our SSS scheme, where shares destined to $i$ are aggregated by its neighbors. (d) shows steps [H6] in our SSS scheme,
    which is equivalent (in term of message flow) to step [H2] in our homomorphic scheme. }\label{dimes}
\end{figure}

\subsection{Extending the method to support multiplication}
\ignore{
In the previous subsections, for simplicity of explanation, we
have resorted to the description of algorithms which perform
additions of messages in each round. Both the homomorphic and the
secret sharing scheme can be extended to support multiplication as
well.
}

Assume  that node $i$ needs to compute the multiplication of the
values of  two messages that it receives from nodes $j$ and $j'$.
\ignore{ Using the Paillier encryption, we use the homomorphic
property of exponentiation, to compute the product of two
encrypted messages. $E_{pk}(m1) \cdot_{pk} E_{pk}(m2) = E_{pk}(m1
\cdot m2)$, where the $\cdot_{pk}$ operation in the Paillier
cryptosystem is exponentiation. } The Shamir secret sharing scheme
can be extended to support multiplication using the construction
of Ben-Or, Goldwasser and Wigderson, whose details appear
in~\cite{BGW}. This requires two changes to the basic protocol.
First, the degree of the polynomials must be strictly less than
$|N_i|/2$, where $|N_i|$ is the number of neighbors of the node receiving
the messages. (This means, in particular, that security is now
only guaranteed as long as less than half of the neighbors
collude.) In addition, the neighboring nodes must exchange a
single round of messages after receiving the messages from nodes
$j$ and $j'$.
We have not implemented this variant of the protocol.

\subsection{Working in different fields}
The operations that can be applied to secrets in the Shamir secret
sharing scheme, or to encrypted values in a homomorphic encryption
scheme, are defined in a finite field or ring over which the schemes
are defined (for example, in the secret sharing case, over a field
$Z_p$ where $p$ is a prime number). The operations that we want to
compute, however, might be defined over the Real numbers.  Working
in a field is sufficient for computing additions or
multiplications of integers, if we know that the size of the field
is larger than the maximum result of the operation. If the basic
elements we work with are Real numbers, we can round them first to
the next integer, or, alternatively, first multiply them by some
constant $c$ (say, $c=10^6$) and then round the result to the
closest integer. (This essentially means that we work with
accuracy of $1/c$ if the computation involves only additions, or
an accuracy of $1/c^d$ if the computation involves summands
composed of up to $d$ multiplications.)

\subsection{Discussion}
\paragraph{Handling division operations.}
Handling division is much harder, since  we are essentially limited
to working with integer numbers. One possible workaround is
possible if we know in advance that a number $x$ might have to be
divided by a different number from a set $D$ (say, the numbers in
the range $[1,100]$). In that case we first multiply $x$ by the
least common multiple (lcm) of the numbers in $D$. This initial
step ensures that dividing the result by a number from $D$ results
in an integer number.

\paragraph{An optimization.}
It is possible to optimize the SSS construction in the case that
the degree of the polynomial used in the vicinity of node $i$
equals the number of node $i$'s neighbors $(d_i - 1 = |N_i|)$. In
this case it is possible to avoid the polynomial evaluation and
interpolation. This is done by replacing steps $[S1,S2]$ with the
following operations: when a node $j$ would like to send a scalar
message $m_{ji}$ to node $i$, it will select $d_i$ numbers at
random, such that their sum in an appropriate finite field is
$m_{ji}$. Steps $[S3-S5]$ remain the same. In step $[S6]$ instead
of node $i$ performing a polynomial interpolation the values it
received, it simply aggregates them using summation to obtain the
weighted sum $\sum_{j \in N_i} a_{ji} m_{ji}$. The drawback of
this method is that there is no redundancy in the received parts,
and as a result even a single neighbor that does not sent its
share to node $i$ can prevent node $i$ from completing its
computation.

\paragraph{Optimization of Lagrange interpolation.}
Each node must compute, in every step of the protocol, a Lagrange
interpolation of the shares it receives. Namely, node $i$ which
receives shares from nodes $u_1,\ldots,u_{d_i}$ must compute the free
coefficient of the corresponding polynomial. This is done by computing
the Lagrange interpolation formula $\sum_{j=1}^{d_i}\lambda_j P(u_j)$,
where $P(u_j)$ is the share received from node $u_j$, and $\lambda_j$
is the corresponding Lagrange coefficient which is defined as
$$\lambda_j = {
  \Pi_{1\leq k \leq d_i; \; k \neq j\; }k \over \Pi_{1\leq k \leq d_i;
    \; k \neq j\; }(k-j)}\;.$$ Note that the computation of
$\lambda_j$ involves many multiplications, but it only depends the
identities of $i$'s neighbors, rather than on the values of their
shares. Therefore, node $i$ can precompute the Lagrange coefficients,
and later use them to compute the linear combination $\sum\lambda_j
P(u_j)$, of the shares it receives. This step considerably reduces the
online overhead of $i$'s operation.

\paragraph{Using a single polynomial for implementing broadcast.}
Assume a setting in which node $j$ needs to broadcast the same value $m_j$
to all its neighbors. (This assumption does not hold in general, however
there are special cases where it does hold, for example the Jacobi algorithm described in \sectionref{Jacobi}.)

In the SSS method described above, node $j$ needs to construct a
different polynomial $P_i$ for each neighbor $i$, encode $m_j$ as the
free coefficient of $P_i$, and send shares of $P_i$ to the neighbors
of node $i$. For implementing broadcast, node $j$ can generate a single polynomial whose
free coefficient is $m_j$, and send values of this polynomial to the
neighbors of each of its neighbors. This is possible if there is an
upper bound of $d^{(j)}-1$ on the number of colluders among the second
degree neighbors of $j$ (i.e., among neighbors of $j$'s neighbors),
and it also holds that no neighbor of $j$ has less than $d^{(j)}$
neighbors. In that case node $j$ sets the degree of the polynomial to
$d^{(j)}-1$, and sends a share of this polynomial to each neighbor of its
neighbor. (Setting the degree to this value enables each neighbor to
interpolate the messages sent to it, yet prevents the colluders from
learning illegitimate information.)

An obvious  advantage of this approach is that $j$
needs to send a single share to each neighbor $u$ of its neighbors,
even if $u$ happens to be a neighbor of two or more of $j$'s
neighbors. In the previous SSS based method, node $j$ needed to send
to $u$ a different share for every node $i$ which is a joint neighbor
of $j$ and $u$.

Consider now  the aggregation operation that is performed by $u$, in which
$u$ computes a linear combination of all shares which are destined to
$i$. These shares are values of different polynomials (generated by
different neighbors of $i$), but the requirement above ensures  that
 each of
these polynomials, say polynomial $P_{j'}$ generated by node $j'$,
is of degree that is
at most $d^{(j')}-1$. This value is smaller than
 $d_i$, the number of neighbors of
$i$. Therefore the linear combination of these polynomials is a
polynomial whose degree is smaller than $d_i$. Node $i$ receives $d_i$
shares of this polynomial and  can therefore interpolate it.

\paragraph{Collusion of  distant nodes in the graph.}
The privacy of the data that node $j$ sends to node $i$, encoded based
on Shamir's secret sharing using a polynomial of degree $d-1$, is
preserved as long as an adversary does not get hold of $d$ shares.
Therefore, if $j$ uses a different polynomial for encoding the
messages sent to each of its neighbors (i.e., it does not use the
method discussed in the previous paragraph), then it only needs to
care about collusions between members of the set of neighbors of each
of its neighbors {\em separately}. (E.g., if $j$ has 8 neighbors, and
it is known that for each neighbor $i$ it holds that no more than 3 of
$i$'s neighbors collude, then $j$ can encode its messages using a
different polynomial of degree 3 for each of its neighbors. This
encoding is secure even if $j$ sends the same message $m_j$ to all its
neighbors, and even if the total number of colluders is much larger
than 3, since each choice of the free coefficients of the polynomials
is plausible given the values known to the colluders.)

If $j$ uses a single polynomial to encode the messages it sends to all
its neighbors, as is detailed in the previous paragraph, then it must
make sure that the degree of the polynomial is at least as large as
the potential number of colluders between all neighbors of its
neighbors. It does not have to care about the integrity of other
members of the network.


\section{Cryptographic background for the malicious adversary model}
In this section, we describe two cryptographic primitives that are used
in our construction for the malicious adversary model. In this section we
give a brief review of those primitives, while in \sectionref{malicious} we
explain how those primitives are used in the context of secure multi-party computation.

\label{crypto2}
\subsection{Pedersen VSS}\label{Pedersen2}

Pedersen~\cite{Pedersen91} presents a non-interactive verifiable
secret sharing scheme (VSS). In this scheme, each party can verify that he received a
correct share, without communicating with any other party.

Pedersen VSS is based on the usage of a commitment scheme which was
also designed by Pedersen. (A commitment scheme enables a committer to
commit to a value without revealing it. Later the committer can reveal
that value. Other parties are assured that the committer was not able
to change the committed value after the commitment was generated.)
The commitment scheme
is based on the assumption that the discrete logarithm problem is hard
in a certain group. The commitment scheme operates in the following
way: Two generators $g$ and $h$ of the group are chosen \ignore{from
  $G_q$ } at random. (The discrete logarithm assumption therefore
implies that computing $\log_g(h)$ is infeasible.)  In order to
commit to a value $s$ \ignore{$\in Z_p^*$}, the committer randomly
chooses a value $t$ \ignore{$\in Z_p^*$} and computes the
commitment $E(s,t) \triangleq g^sh^t$.
Then, in order to open the commitment, the committer reveals $s$
and $t$.

It was proven that the committer cannot change her mind after
generating the commitment. This was proved by showing that a committer
that can change the committed value from $s$ to a different value $s'$
can compute $\log_g(h)$. Pedersen then showed how to use this primitive
to share a secret $s$ between $n$ parties in a way that enables them
to verify that their shares are consistent. We describe this protocol
below,
where the
dealer $D$ plays the role of the committer of the basic commitment
scheme.

\begin{itemize}
\item [{[VS1]}] $D$ performs the basic commitment scheme and computes a
  commitment of the secret $s$: $E_0 \triangleq E(s,t)\,$.
\item [{[VS2]}] $D$ performs the first step of Shamir secret sharing
  scheme: $D$ randomly chooses a polynomial $P(x)$ of degree at most
  $d-1$ subject to the constraint $P(0) = s$ and computes $P(i)$ for $i=1,...,n$ (we
  denote $P(x)=\sum_{k=0}^{d-1}p_kx^k$; therefore $p_0=s$).
\item [{[VS3]}] $D$ randomly chooses a polynomial $R(x)$ of degree at
  most $d-1$  subject to the constraint
 $R(0) = t$ and computes $R(i)$ for $i=1,...,n$ (we
  denote $R(x)=\sum_{k=0}^{d-1}r_kx^k$; therefore $r_0=t\,$.)
\item [{[VS4]}] $D$ performs the second step of Shamir secret sharing
  scheme: $D$ secretly sends $(P(i), R(i))$, the i'th share, to party
  $i$, for $i=1,...,n$.
\item [{[VS5]}] $D$ computes and broadcasts a commitment to $P(x)$'s
  coefficients
  $p_0,...,p_{d-1}$ using $R(x)$'s coefficients
  $r_0,...,r_{d-1}$. I.e.,  $D$
  broadcasts $E_j \triangleq E(p_j,r_j)$ for $j=0,...d-1$.
 Denote $\mathbf{E} = (E_0, E_2, \cdots, E_{d-1})$ the set of all
 commitments.
\item [{[VS6]}] Party $i$ can now  verify that the  share $(P(i), R(i))$
  that
  it received is correct. This is done by verifying the equation
$E(P(i), R(i)) = \prod_{j=0}^{d-1}(E_j)^{i^j}$.

\end{itemize}
Note that when the parties send their shares to the party who is
supposed to combine them in order to compute the secret, the
verification  data $E_0,\ldots,E_{d-1}$ can  be used
to verify the correctness of the received shares.

\subsection{Byzantine agreement}
\label{agreement}

The Byzantine agreement (Byzantine Generals)
problem was first introduced by Pease, Shostak and Lamport
\cite{Agree80}. It is now considered as a fundamental problem in
fault-tolerant distributed computing. The task is to reach agreement
in a set of $n$ nodes in which up-to $f$ nodes may be faulty. A
distinguished node (\emph{the General} or \emph{the initiator})
broadcasts a value \emph{m}, following which all nodes exchange
messages until the non-faulty nodes agree upon the same value. If
the initiator is non-faulty then all non-faulty nodes are required
to agree on the same value that the initiator sent.

On-going faults whose nature is not predictable or that
express complex behavior are most suitably addressed in the
Byzantine fault model. It is the preferred fault model in order to
seal off unexpected behavior within limitations on the number of
concurrent faults. With respect to the bounds on redundancy, the
Byzantine agreement problem has been shown to have no deterministic
solution if more than $n/3$ of the nodes are concurrently faulty
\cite{Agree82}.

A Byzantine Agreement protocol satisfies the following typical
properties:\\

\noindent{\bf Agreement:} The protocol returns the same value at all
correct nodes;

\noindent{\bf Validity:} If the General is correct,  then  all the
correct nodes return the value sent by the General;

\noindent{\bf Termination:} The protocol terminates in a finite
time.\\

Standard deterministic Byzantine agreement algorithms operate in the
synchronous network model in which it is assumed that all correct
nodes initialize the agreement procedure and can exchange messages
within a round.  In the context of our paper one can use the modular
solution appearing in~\cite{FastAgree87}. That protocol requires
$2f+1$ rounds of communication and $O(nf^2)$ messages, where $f$ is
the bound on the number of malicious nodes. The protocol can be
invoked by each node that wants to broadcast a message. The set of
participating nodes are the General's direct neighbors. In some cases
we may define a larger set, a set that contains some neighbors of
neighbors.
It is assumed that every pair of nodes in the set can exchange messages.
For that to hold we assume that  the network connectivity among the members of the set will be at least $2f+1$ (see~\cite{d82}).

Byzantine agreements are guaranteed to provide the same value at all participating non-faulty nodes. If the general is faulty that value may turn out to be some default value.  In the context of our paper, since we carry out computations based of the values  sent by all the neighbors of a node we assume that the default value is a zero. In the context of the current paper we also assume that for each node $i$, $d_i>f.$

\section{Extending the construction to support malicious adversaries}
\label{malicious}

In this section we extend the SSS protocol of \sectionref{our_SSS}, to
defend against malicious peers.  The new protocol utilizes the
mechanisms described in the previous section.  Before presenting the
full protocol, we devise a modified VSS scheme.  This scheme is
needed, since the original VSS relies on a broadcast primitive, which
does not exist in a Peer-to-Peer network. The modified protocol is
following Pedersen's VSS scheme, except of step [VS5] which is
replaced by a Byzantine agreement. (Byzantine agreement is used in
order to ensure that all relevant nodes receive the same verification
information, i.e. exponents of coefficients, of Pedersen's protocol.)

\subsection{Modified VSS}
In each round of the algorithm, when a node $j$ would like to send
a scalar value $m_{ji}$ to node $i$ it does the following operations:
\begin{itemize}
\item [{[MV1]}] Node $j$ perform steps [VS1-VS4] in Pedersen's
scheme for creating verifiable shares of the message $m_{ji}$ and
sends the shares to node $i$'s neighbors.

\item [{[MV2]}] Node $j$ runs a Byzantine agreement protocol
  between $j$ and all of node $i$'s neighbors (including $i$ itself), in which node $j$
  broadcasts the set $\mathbf E$ of commitments to the coefficients of
  the polynomials. (We denote
  this set as $\mathbf E^{(ji)}$ later in the protocol.)  This step
  replaces the broadcast primitive in [VS5] that does not exist
  in a Peer-to-Peer network.

\item [{[MV3]}] Node $i$'s neighbors verify the validity of the shares
using [VS6]. In case the share received by neighbor $l$ is not valid,
this neighbor informs node $i$ about that and does not send it the
required linear combination of shares.
\end{itemize}


\subsection{The full protocol}

Initialization:
\begin{itemize}
\item [{[BS0]}] We assume that the coefficients $a_{ji}$ are known in
  advance to all neighbors of $i$. (If that is not the case, then
  these coefficients are decided by either node $i$ or $j$. That node
  must perform with node $i$'s neighbors a Byzantine agreement for the
  value $a_{ji}$.)
\end{itemize}
This step is needed because node $i$'s neighbors will verify the weighted
sum computation it will carry out later.\\
\\
In each round of the algorithm, when a node $j$ would like to send
a scalar value $m_{ji}$ to node $i$ it does the following operations:
\begin{itemize}
\item [{[BS1]}] Protocol [MV1-3] is executed for sending and verifying
  the shares of the message $m_{ji}$. (This step is only executed in
  the first round of the protocol. In later rounds it is replaced by
  Step [BS6] in which the neighbors also verify that  $m_{ji}$ is
  indeed the message that $j$ was supposed to send according to the
  protocol.)

\item [{[BS2]}] Each neighbor $l$ of node $i$ that validated all the
  shares of all the neighbors of node $i$ aggregates the shares it
  received from all these neighbors using linear coefficients $\{
  a_{ji}\; | \; j\in N_i\}$, and computes the value $S_{li} = \sum_{j
    \in N_i} a_{ji} P_{ji}(l)$, and the value $T_{li} = \sum_{j \in
    N_i} a_{ji} R_{ji}(l)$.  (This computation computes the values of
  two polynomials of degree $d_i-1$, whose free coefficients are equal
  to the weighted sum of all messages $m_{ji}$ sent to node $i$
  by its neighbors, and of all values $t_{ji}$ that are used for
  computing the corresponding commitments.)

\item [{[BS3]}] Each neighbor $l$ sends the sums $S_{li}, T_{li}$
      to node $i$.

\item [{[BS4]}] Node $i$ first verifies the values it received
      from its neighbors: Each neighbor must
      essentially send values of polynomials generated as linear
      combinations of the polynomials $P_{ji},R_{ji}$ of all neighbors
      $j$ of $i$, for which
      commitments has been sent. Therefore a linear combination of the
      commitments can be used to verify the values received from the
      neighbors.  In more detail, let $\mathbf{E^{(ji)}}=
      (E_0^{(ji)},\ldots,E_{d-1}^{(ji)})$ be the commitments
      sent by node $j$ with respect to the message
      $m_{ji}$. For every neighbor $l$ which sent to $i$ the values $S_{li},
      T_{li}$, node $i$ verifies that $E(S_{li}, T_{li}) =
      \Pi_{k=0}^{d-1} (E_k^{(i)})^{l^k}$, where $E_k^{(i)}$ is equal
      to $\Pi_{j \in N_i} (E_k^{(ji)})^{a_{ji}}$.

\item [{[BS5]}] If $d_i$ or more messages sent from neighbors were verified
      correctly, node $i$ considers these $S_{li}$ values as values of a
      polynomial $P_i$ of degree $d_i-1$. It interpolates $P_i(x)$ in
      order to extract the free coefficient, which in this case is the
      weighted sum of messages sent to it, $\sum_{j \in N_i} a_{ji}
      m_{ji}$.

    \item [{[BS6]}] Note that the weighted sum $m'=\sum_{j \in N_i}
      a_{ji} m_{ji}$ which was calculated by $i$ is included in the message
      that $i$ must send to its neighbors in the next round of the
      protocol. (Namely, $i$ must send to every neighbor $i'$ of his
      the message $m_{ii'}=a_{ii'} m'$.) In addition, we will require that the
      $t'$ value that $i$ uses in a VSS for computing $E'_0=g^{m'}h^{t'}$
      will be $\sum_{j \in N_i} a_{ji} t_{ji}$, namely be equal to the
      linear combination of the $t$ values in the messages sent to $i$.

      The neighbors of $i$ must therefore verify that they receive
      shares of $m'$.  This is done in the following way: Each
      neighbor $i'$ first computes a linear combination of the free
      coefficient of the Pedersen commitments (of messages destined to
      $i$) that it received in the last
      step: $E_0^{(i)} = \Pi_{j \in N_i}
      (E_0^{(ji)})^{a_{ji}}\,\,$. Node $i$ runs a modified VSS
      protocol where it commits to $m'$ using the value $t'$ and
      polynomials $P'()$ and $R'()$ of the required degrees. The result is a
      vector $\mathbf{E'}$\ whose first entry is $E^{(i)}_0$ as defined
      above.  Node $i$ runs steps [MV1-MV3] with these values,
      sending them to all its neighbors. Each of the neighbors
      verifies that the first entry of $\mathbf{E'}$\  is indeed
      $E^{(i)}_0$.
Later in the execution of  this step,
      each node $i'$ verifies the linear combination it receives from its
      neighbors  using  the values  $\{ E^{(i)}_0 \; | \; i\in N_{i'} \}$.
If a verification by a node fails, it
      aborts the protocol and notifies its neighbors.


\end{itemize}

\subsection{Protocol analysis}
\label{prot_analysis}

The extended protocol complexity is higher than the SSS protocol
for the semi-honest model. Below we present an
analysis of the efficiency of the extended protocol in terms of
computational and message overhead.

First we list the computational and message overhead of the
building blocks of the protocol, and then we sum them.

\begin{itemize}
\item \textbf{Byzantine Agreement:} A Byzantine agreement of
$|N_i|$ nodes of which at most $d$ nodes are malicious consists of
$|N_i|d^2$ messages in $2d+3$ communication rounds.

\item \textbf{Polynomial Creation:} Creating a random polynomial
of degree $d-1$ costs $O(d)$ random number generation operations;
this computational overhead is negligible in the overall
computational overhead.

\item \textbf{Polynomial Evaluation:} Evaluating of a polynomial
of degree $d-1$ costs $O(d^2)$ multiplication operations.

\item \textbf{Values Verification:} Verifying of a value (a share
or a weighted sum) costs $O(d)$ exponentiation operations.

\item \textbf{Polynomial Interpolation:} Interpolating a polynomial of
  degree $d-1$ costs $O(d^2)$ multiplication operations.

\end{itemize}

\noindent \textbf{Message overhead:}~ The dominant element
regarding message overhead is the Byzantine agreement protocol
(Steps [BS0], [BS1], [BS6])
 that requires $|N_i|d^2$ messages in
$2d+3$
communication rounds.\\

\noindent \textbf{Computational overhead}~ The dominant element
regarding computational overhead is the values verifications
(Steps [BS1],
[BS4], [BS6])
that all together cost $O(|N_i|d)$ exponentiation operations.\\

There are some means to minimize the number of computing
operations:

\begin{itemize}
\item \textbf{Polynomial evaluation optimization.} Polynomial
evaluation, that normally takes $\frac{(d-1)d}{2}$
multiplications, can be optimize to take $d$ multiplications, if
the value of the input parameter is bounded in a known range and
all of $d$ exponentiations of all the possible values are prepared
ahead, e.g., in the initialization step (this implicitly requires
knowing $d$ ahead). This decreases the number of computing
operations in Steps [BS1]
and [BS2]. A similar idea can be implemented in the verification
in Steps [BS1],
[BS4], and [BS6].

\item \textbf{Commitments calculations optimization.} The
commitments that are agreed on Step [MV2] $\mathbf{E^{(ji)}}=
(E_0^{(ji)},\ldots,E_{d-1}^{(ji)})$ can be computed ahead if the
polynomials coefficients are bounded in a known range. This can be
done by preparing ahead commitments $E_{x,y} = E(x,y) = g^xh^y$
for each $x,y$ that are in the bounded range. This decreases the
number of computing operations in Step [MV2].

\end{itemize}

\paragraph{Security.} We argue here that either every party follows
the protocol, or the protocol aborts. Every modified VSS protocol uses
Byzantine agreement to broadcast the verification data
$\mathbf{E}$. Therefore either all honest nodes receive shares
corresponding to the same polynomial, or the protocol
aborts. Furthermore, the linear combinations of these shares that are
sent to node $i$ can be verified by the same data, and therefore no
neighbor of $i$ can send a corrupt linear combination.
Finally, in the next step of the protocol node $i$ must send a linear
combination of the messages it received. This is verified by its
neighbors, by using the same verification data, as is detailed in
Step~[BS6] of the protocol.

For using the Byzantine agreement protocol, we demand that $f$, the
number of malicious peer in each vicinity is less than $1/3$ of the peers
in that vicinity. Alternatively, a Byzantine agreement with signatures
can be used, tolerating any number of malicious nodes.


\subsection{Discussion}
\label{insensitive}
\paragraph{A more realistic model.}
The Shamir Secret Sharing protocol makes sure that one party's share
is not revealed by other parties, unless at least $t$ parties
cooperate ($t-1$ is the degree of the polynomial). However, a model on
which most of the parties are insensitive to their privacy is
realistic. In such a model, there is no reason to make an effort in
order to prevent revealing shares of these insensitive nodes. This
observation can be refined by setting a ``paranoic coefficient'' for
each node that describes the extent of privacy-sensitivity of this
node. As the ``paranoic coefficient'' decreases, so does the degree of
the polynomial. In particular, insensitive nodes can use a polynomial
of degree $0$.

\ignore{
\paragraph{An optimization.}
\bp{I'll modify this paragraph}
The usage of a Byzantine agreement
primitive complicates the use of
Pedersen's VSS in the case of malicious adversaries. The protocol
can be simplified if we only want to make sure that node $v$ can
verify that all the values it receives from its neighbors are
correct. Each neighbor $u$ of $v$ will send shares to every other
neighbor $u'$ of $v$, according to a polynomial $P()$. It will
also send $g$ raised to the power of coefficients of $P()$ to $v$
alone. This is insufficient in order for $u'$ to verify that the
share it receives from $u$ is correct. However, when $v$ receives
from $u'$ a linear combination of the shares received by $u'$, it
can use the verification data that it previously received to
verify that it got a correct combination of the shares (recall
that $v$ knows the coefficients of the linear combination). If the
verification fails then $v$ knows that someone cheated, but it
cannot identify who it was. \db{Benny's comment: The arguments
here should be proved but I think that they are correct.}
}

\paragraph{Synchronous vs. Asynchronous execution.}
In this paper, we have presented the iterative algorithm that computes
weighted sums as a synchronous algorithm which operates in rounds.
This was mainly done for simplifying our exposition.
However, in practice it is not  valid to assume  that the clocks and message delays are synchronized in
a large Peer-to-Peer network. Luckily, it is known that linear
iterative algorithms such as the Jacobi algorithm
converge in asynchronous settings as well. Specifically, the Gauss Seidel algorithm
is an asynchronous version of the Jacobi algorithm which typically converges faster~\cite{BibDB:BookBertsekasTsitsiklis}.

\paragraph{An optimization to the Byzantine agreement protocol.}
It is possible to optimize the Byzantine agreement by using a Public Key infrastructure that enables signatures.
The existence of the Public Key infrastructure limits the ability of the malicious nodes to introduce undetected superfluous messages.  Algorithms that reach Byzantine agreement under such assumptions require sending only a constant number of messages of each node to all participating nodes~\cite{DS82}. Moreover, such protocols can overcome any ratio of faulty to correct nodes.

Another optimization is to replace the deterministic protocol with a probabilistic one (cf.~\cite{Benor83,FM89,CR93}). A typical probabilistic protocol terminates in an expected constant number of rounds.  The drawback is that such protocols do not guarantee that all non-faulty nodes complete the protocol at the same time. In our context this implies that a node waits a round before sending messages based on the agreement, to ensure that others also completed the protocol. Our protocol requires running several agreements in concurrently.  The asynchronous nature of the execution allow nodes to use each value once the agreement about it is completed.

\section{Case Study: neighborhood based collaborative filtering}
\label{jacobi} To demonstrate the usefulness of our approach, we
give a specific instance of a problem our framework can solve,
preserving users' privacy. Our chosen example is in the field of
collaborative filtering. We have chosen to implement the
neighborhood based collaborative filtering algorithm, a
state-of-the-art algorithm, winner of the Netflix progress prize
of 2007. When adapting this algorithm to a Peer-to-Peer network,
there are two main challenges: first, the algorithm is
centralized, while we would like to distribute it, without losing
accuracy of the computed result. Second, we would like to add a
privacy preserving layer, which prevents the computing nodes from
learning any information about neighboring nodes or other nodes
rating, except of the computed solution.

We first describe the centralized version, and later we extend it
to be computed in a Peer-to-Peer network. Given a possibly sparse
user ratings matrix $\mR_{m \times n}$, where $m$ is the number of
users and $n$ is the number of items, each user likes to compute
an output ratings for all the items.

In the neighborhood based approach~\cite{KorenCF}, the output
rating is computed using a weighted average of the neighboring
peers:
\[ r_{ui} = \sum_{j \in N_i} r_{uj} w_{ij}. \]
Our goal is to find the weights matrix $\mW$ where $w_{ij}$
signifies the weight node $i$ assigns node $j$.

We define the following least square minimization problem for user
$i$ :
\[ \min_\vw \sum_{v \ne i}(r_{vi} - \sum_{j \in N_i}
r_{vj} w_{ij})^2\;. \]

The optimal solution is formed by differentiation and solution of
a linear systems of equations $\mR \vw = \vb$. The optimal weights
(for each user) are given by: \BE \label{eqw} \vw = (\mR^T
\mR)^{-1} \mR^T \vb\;. \EE

We would like to distribute the neighborhood based collaborative
filtering problem to be computed in a Peer-to-Peer network. Each
peer has its own rating as input (the matching row of the matrix
$\mR$) and the goal is to compute locally, using interaction with
neighboring nodes, the weight matrix $\mW$, where each node has
the matching row in this matrix. Furthermore, the peers would like
to keep their input rating private, where no information is leaked
during the computation to neighboring or other nodes. The peers
will obtain only their matching output rating as a result of this
computation.

We propose a secure multi-party computation framework, to solve the
collaborative filtering problem efficiently and distributively,
preserving users' privacy. The computation does not reveal any
information about users' prior ratings, nor on the computed
results.

\subsection{The Jacobi algorithm for solving systems of
linear equations}\label{Jacobi} In this section we give an example
of one of the simplest iterative algorithms for solving systems of
linear equations, the Jacobi algorithm. This will serve as an
example for an algorithm our framework is able to compute, for
solving the neighborhood based collaborative filtering problem.
Note that there are numerous numerical methods we can compute
securely using our framework, among them Gauss Seidel, EM
(expectation minimization), Conjugate gradient, gradient descent,
Belief Propagation, Cholskey decomposition, principal component
analysis, SVD etc.

Given a system of linear equations $\mA \vx = \vb$, where $\mA$ is
a matrix of size $n \times n$, $\forall_i a_{ii} \ne 0$ and $\vb
\in \mathbb{R}^n$, the Jacobi
algorithm~\cite{BibDB:BookBertsekasTsitsiklis} starts from an
initial guess $\vx^{0}$, and iterates: \BE \label{Jeq}
 x_i^{r} = \frac{b_i - \sum_{j \in N_i}a_{ij} x_j^{r-1}}{ a_{ii}}\;.
 \EE
The Jacobi algorithm is easily distributed since initially each
node selects an initial guess $x^{0}_i$, and the values $x_j^r$
are sent among neighbors. A sufficient condition for the algorithm
convergence is when the spectral radius
 $\rho(I - D^{-1}\mA) < 1$,
where $I$ is the identity matrix and $D = \mbox{diag}(\mA)$. This
algorithm is known to work in asynchronous settings as well. In
practice, when converging, the Jacobi algorithm convergence speed
is logarithmic\footnote{Computing the pseudo inverse
solution (equation \ref{eqw}) iteratively can be done more efficiently
using newer algorithms, for example~\cite{ISIT2}. For the purpose
of the clarify of explanation, we use the Jacobi algorithm. } in $n$.

 Our goal is to compute a {\em privacy-preserving}
version of the Jacobi algorithm, where the inputs of the nodes are
private, and no information is leaked during the rounds of the
computation.

Note, that the Jacobi algorithm serves as an excellent example
since its simple update rule contains all the basic operation we
would like to support: addition, multiplication and substraction.
Our framework supports all of those numerical operations, thus
capturing numerous numerical algorithms.

\subsection{Using the Jacobi algorithm for solving the
neighborhood based collaborative filtering problem} First, we
perform a distributed preconditioning of the matrix $\mR$. Each
node $i$ divides its input row of the matrix $\mR$ by $R_{ii}$.
This simple operation is done to avoid the division in \eqnref{Jeq},
while not affecting the solution vector $\vw$.


Second, since Jacobi algorithm's input is a square $n \times n$
matrix, and our rating matrix $\mR$ is of size $m \times n$, we
use the following ``trick'': We construct a new symmetric data
matrix $\tilde{\mR}$ based on the non-rectangular rating matrix
$\mR\in\mathbb{R}^{m\times n}$ \BE \label{newR}
\tilde{\mR}\triangleq\left(
  \begin{array}{cc}
    \mI_{m} & \mR^T \\
    \mR & 0 \\
  \end{array}
\right)\in\mathbb{R}^{(m+n)\times(m+n)}\;. \EE Additionally, we
define a new vector of variables
$\tilde{\vw}\triangleq\{\hat{\vw}^{T},\vz^{T}\}^{T}\in\mathbb{R}^{(m+n)\times1}$,
where $\hat{\vx}\in\mathbb{R}^{m\times1}$ is the (to be shown)
solution vector and $\vz\in\mathbb{R}^{n\times1}$ is an auxiliary
hidden vector, and a new observation vector
$\tilde{\vb}\triangleq\{\mathbf{0}^{T},\vb^{T}\}^{T}\in\mathbb{R}^{(m+n)\times1}$.

Now, we would like to show that solving the symmetric linear
system $\tilde{\mR}\tilde{\vw}=\tilde{\vb}$, taking the first $m$
entries of the corresponding solution vector $\tilde{\vw}$ is
equivalent to solving the original system $\mR\vw=\vb$. Note that
in the new construction the matrix $\tilde{\mR}$ is still sparse,
and has at most $2mn$ off-diagonal nonzero elements. Thus, when
running the Jacobi algorithm we have at most $2mn$ messages per
round.

Writing explicitly the symmetric linear system's equations, we get
\[ \hat{\vw}+\mR^T\vz=\mathbf{0},\mbox{  }\\
    \mR\hat{\vw}=\vb.
    \]

By extracting $\hat{\vw}$ we obtain \[
\hat{\vw}=(\mR^{T}\mR)^{-1}\mR^{T}\vb, \] the desired solution of
\eqnref{eqw}.

\section{Experimental Results}
\label{exp} We have implemented our proposed constructions
for the semi-honest model using a
large scale simulation. Our simulation is written in C, consists
of about 1500 lines of code, and uses MPI, for running the
simulation in parallel. We run the simulation on a cluster of
Linux Pentium IV computers, 2.4Ghz, with 4GB RAM memory. We use
the open source Paillier implementation of~\cite{PaillierIMP}.
Currently, we have implemented fully the semi-honest protocols.
An area of future work is to implement the full protocol against
the malicious participants as well.

\begin{figure}
  \includegraphics[width=200pt]{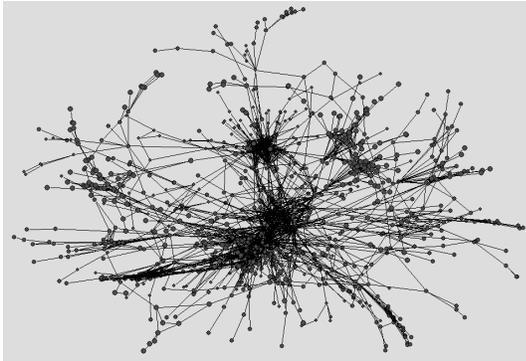}\\
  \caption{DIMES Internet router topology consisting around 300K routers and 2.2M communication links. A subgraph containing 500 nodes is shown.}\label{fig:DIMES}
\end{figure}

We use several large topologies for demonstrating the
applicability of our approach. The different topologies are listed
in Table~\ref{tb1}. The DIMES dataset \cite{DIMES} is an Internet
router topology of around 300,000 routers and 2.2 million
communication links connecting them, captured in January 2007. A
subgraph of the DIMES dataset is shown in Figure~\ref{fig:DIMES}.
The Blog network, is a social network, web crawl of Internet blogs
of half a million blog sites and eleven million links connecting
them. Finally, the Netflix~\cite{Netflix} movie ratings data,
consists of around 500,000 users and 100,000,000 movie ratings.
This last topology is a bipartite graph with users at one side,
and movies at the other. This topology is not a Peer-to-Peer
network, but relevant for the collaborative filtering problem. We
have artificially created a Peer-to-Peer network, where each user
is a node, the movies are nodes as well, and edges are the ratings
assigned to the movies.

\begin{table}[h!]
\begin{center}
\begin{tabular}{|c|c|c|c|}
  \hline
  Topology & Nodes & Edges & Data Source \\
  \hline
  Blogs Web Crawl & 1.5M & 8M & IBM \\
  DIMES & 337,326  & 2,249,832 & DIMES  \\
  Netflix & 497,759 & 100M & Netflix \\
  \hline
\end{tabular}
\caption{\mbox{              } Topologies used for
experimentation}
\end{center}
\end{table}\label{tb1}
\vspace{-5mm}

We ignore algorithm accuracy since this problem was addressed in
detail in~\cite{KorenCF}. We are mainly concerned with the
overheads of the privacy preserving mechanisms. Based on the
experimental results shown below, we conclude that the main
overhead in implementing our proposed mechanisms is the
computational overhead, since the communication latency exists
anyway in the underlying topology, and we compare the run of
algorithms with and without the added privacy mechanisms overhead.
For that purpose, we ignore the communication latency in our
simulations. This can be justified, because in the random
perturbations and homomorphic encryption schemes, we do not change
the number of communication rounds, so the communication latency
remains the same with or without the added privacy preserving
mechanisms. In the SSS scheme, we double the number of
communication rounds, so the incurred latency is doubled as well.

\tableref{tb2} compares the running times of the basic operations in the
three schemes. Each operation was repeated 100,000 times and an
average is given. As expected the heaviest computation is the
Paillier asymmetric encryption, with a security parameter of 2,048
bits. It can be easily verified, that while the SSS basic
operation takes around tens of microseconds, the Paillier basic
operations takes fractions of seconds (except of the homomorphic
multiplication which is quite efficient since it does not involve
exponentiation). In a Peer-to-Peer network, when a peer has likely
tens of connections, sending encrypted message to all of them will
take several seconds. Furthermore, this time estimation assumes
that the values sent by the function are scalars. In the vector
case, the operation will be much slower.

\tableref{tb3} outlines the running time needed to run 8 iterations of
the Jacobi algorithm, on the different topologies. Four modes of
operations are listed: no privacy preserving means we run the
algorithm without adding any privacy layer for baseline timing
comparison. Next, our three proposed schemes are shown.

In the Netflix dataset, we had to use eight computing nodes in
parallel, because our simulation memory requirement could not fit
into one processor.

As clearly shown in \tableref{tb3}, our SSS scheme has significantly
reduced computation overhead relative to the homomorphic
encryption scheme, while having an equivalent level of security
(assuming that the Paillier encryption is semantically secure). In
a Peer-to-Peer network, with tens of neighbors, the homomorphic
encryption scheme incurs a high overhead on the computing nodes.

\begin{table*}[t!]
  \centering
\begin{tabular}{|l|l|c|c|}
  \hline
  Scheme & Operation & Time (micro second) & Msg size (bytes)\\ \hline
  Random perturbation & Adding noise & 0.0783745  & 8\\
                     & Receiver operation & $-$ &\\ \hline
  SSS & Polynomial generation and evaluation & 11.18382125 & 8 \\
          & Polynomial extrapolation & 6.13709025 & \\ \hline
  Paillier & Key generation & 5016199.4 & 2048\\
      & Encryption & 203478.62  &\\
 & Decryption & 193537.97   &\\
        & Multiplication & 99.063958 & \\
   \hline
\end{tabular}
  \label{local}
  \caption{Running time of local operations. As expected, the Paillier cryptosystem basic operations
  are time consuming relative to the SSS scheme.}\label{tb2}
\end{table*}

\begin{table*}
  \centering
\begin{tabular}{|l|l|c|c|}
  \hline
  Topology & Scheme & Time (HH:MM:SS) & computing nodes\\ \hline
  DIMES & None & 0:33.36 & 1\\
  & Random Perturbations & 0:35.27 & 1\\
         & SSS & 10:53.44 & 1\\
          & Paillier & 28:44:24.00 & 1\\ \hline
  Blogs & None & 1:28.16 & 1\\
        & Random Perturbations & 1:34.85 & 1\\
         & SSS & 38:00.24 & 1\\
          & Paillier & 101:52:00.00 & 1\\ \hline
  Netflix & None & 5:31.14 & 8\\
  & Random Perturbations & 5:54.69 & 8\\
         & SSS & 21:40.00 & 8 \\
          & Paillier & - & -\\
   \hline
\end{tabular}
  \caption{Running time of eight iterations of the Jacobi algorithm. The baseline timing is compared to running without
  any privacy preserving mechanisms added. Empirical results show that computation time of the homomorphic scheme is a factor of about 1,350 times slower
  then the SSS scheme.}\label{tb3}
\end{table*}

\section{Conclusion and Future Work}
\label{Conclusion} As demonstrated by the experimental results
section, we have shown that the secret sharing scheme in the
semi-honest model has the lowest computation overhead relative to
the other schemes. Furthermore, this scheme does not involve a
trusted third party, as needed by the homomorphic encryption
scheme for the threshold key generation phase. The size of the
messages sent using this method is about the same as in the
original method, unlike the homomorphic encryption which
significantly increases message sizes. However, the  drawback of
this scheme is that neighboring nodes to node $i$ need to
communicate directly between themselves (and each message sent to
node $i$ needs to be converted to messages sent to all its
neighbors). In Peer-to-Peer systems with locality property it
might be reasonable to assume that communication between the
neighbors of node $i$ is possible. (There is a way to circumvent
this requirement, by adding asymmetric encryption. Each node will
have a public key, where message destined to this node are
encrypted using its public key. That way if node $j$ needs to send
a message to node $l$, it can ask node $i$ do deliver it, while
ensuring that node $i$ does not learn the  content of the message.
We identify this extension to our scheme as an area for future
work.)

In the current work, we have extended the SSS protocol to defend against
malicious participants. We have shown that the extension provides a completely secure solution.
However, the main drawback is the high protocol overhead, since we need to perform multiple
Byzantine agreement protocols and commitments, for verifying every single computation done in the network.
An area of future work is to bridge between our theoretical work for the malicious case to
a practical deployment in a Peer-to-Peer network.

Another area of future work is the extension of the homomorphic
protocol to support malicious participants. One possible approach
is to utilize the threshold Paillier cryptosystem supports
verification keys~\cite{Paillier2}, that enables participants  to
verify validity of encrypted messages.

\section*{Acknowledgement}
We would like to thank Dr. Adam Wierzbicki for useful discussions and his helpful comments, especially regarding
realistic models of privacy, where some of the nodes do not care about exposing their inputs.

\bibliographystyle{plain}
\bibliography{PPNA09}

\begin{thebibliography}{10}

\bibitem{GMP}
The {GNU MP Bignum} library. {\tt http://gmplib.org}.

\bibitem{Netflix}
{Netflix. } {\tt www.netflix.org }.

\bibitem{PaillierIMP}
Paillier {C} implementation by {John Bethencourt}. {\tt
  http://acsc.csl.sri.com/libpaillier/}.

\bibitem{AS}
Rakesh Agrawal and Ramakrishnan Srikant.
\newblock Privacy-preserving data mining.
\newblock In {\em Proceedings of the 2000 ACM SIGMOD International Conference
  on Management of Data, May 16-18, 2000, Dallas, Texas, USA}, pages 439--450.
  ACM, 2000.

\bibitem{EWSN08}
Tal Anker, Danny Bickson, Danny Dolev, and Bracha Hod.
\newblock Efficient clustering for improving network performance in wireless
  sensor networks.
\newblock {\em In European Conference on Wireless Sensor Networks (EWSN'08)},
  2008.

\bibitem{KorenCF}
R.~M. Bell and Y.~Koren.
\newblock Scalable collaborative filtering with jointly derived neighborhood
  interpolation weights.
\newblock In {\em IEEE International Conference on Data Mining (ICDM'07)},
  2007.

\bibitem{FairPlayMP}
A.~Ben-David, N.~Nisan, and B.~Pinkas.
\newblock Fairplaymp -- a system for secure multi-party computation.
\newblock manuscript, 2008.

\bibitem{Benor83}
M.~Ben-Or.
\newblock Another advantage of free choice (extended abstract): Completely
  asynchronous agreement protocols.
\newblock In {\em PODC '83: Proceedings of the second annual ACM symposium on
  Principles of distributed computing}, pages 27--30, 1983.

\bibitem{BGW}
M.~Ben-Or, S.~Goldwasser, and A.~Wigderson.
\newblock Completeness theorems for non-cryptographic fault-tolerant
  distributed computation.
\newblock In {\em 20th STOC}, pages 1--10, 1988.

\bibitem{BibDB:BookBertsekasTsitsiklis}
D.~P. Bertsekas and J.~N. Tsitsiklis.
\newblock {\em Parallel and Distributed Calculation. Numerical Methods.}
\newblock Prentice Hall, 1989.

\bibitem{ISIT2}
D.~Bickson, O.~Shental, P.~H. Siegel, J.~K. Wolf, and D.~Dolev.
\newblock Gaussian belief propagation based multiuser detection.
\newblock In {\em IEEE Int. Symp. on Inform. Theory (ISIT), Toronto, Canada,
  July 2008, to appear.}

\bibitem{p2p-rating}
Danny Bickson, Dahlia Malkhi, and Lidong Zhou.
\newblock Peer to peer rating.
\newblock {\em In the 7th IEEE Peer-to-Peer Computing}, 9 2007.

\bibitem{CR93}
R.~Canetti and T.~Rabin.
\newblock Fast asynchronous byzantine agreement with optimal resilience.
\newblock In {\em 25th STOC, Proceedings of the twenty-fifth annual ACM
  symposium on Theory of computing}, 1993.

\bibitem{Canny}
John Canny.
\newblock Collaborative filtering with privacy via factor analysis.
\newblock In {\em SIGIR '02: Proceedings of the 25th annual international ACM
  SIGIR conference on Research and development in information retrieval}, pages
  238--245, New York, NY, USA, 2002. ACM.

\bibitem{DiNi}
Irit Dinur and Kobbi Nissim.
\newblock Revealing information while preserving privacy.
\newblock In {\em PODS '03: Proceedings of the twenty-second ACM
  SIGMOD-SIGACT-SIGART symposium on Principles of database systems}, pages
  202--210, New York, NY, USA, 2003. ACM.

\bibitem{d82}
D.~Dolev.
\newblock The byzantine generals strike again.
\newblock {\em Journal of Algorithms}, 3:14--30, 1982.

\bibitem{DS82}
D.~Dolev and R.~H. Strong.
\newblock Polynomial algorithms for multiple processor agreement.
\newblock In {\em 14th STOC, Proceedings of the twenty-fifth annual ACM
  symposium on Theory of computing}, 1982.

\bibitem{RandomNoise}
Haimonti Dutta, Hillol Kargupta, Souptik Datta, and Krishnamoorthy Sivakumar.
\newblock Analysis of privacy preserving random perturbation techniques:
  further explorations.
\newblock In {\em WPES '03: Proceedings of the 2003 ACM workshop on Privacy in
  the electronic society}, pages 31--38, New York, NY, USA, 2003. ACM.

\bibitem{FM89}
P.~Feldman and S.~Micali.
\newblock An optimal probabilistic algorithm for synchronous byzantine
  agreement.
\newblock In {\em ICALP '89: Proceedings of the 16th International Colloquium
  on Automata, Languages and Programming}, pages 341--378, 1989.

\bibitem{Paillier2}
Pierre-Alain Fouque, Guillaume Poupard, and Jacques Stern.
\newblock Sharing decryption in the context of voting or lotteries.
\newblock In {\em Financial Cryptography, volume 1962 of Lecture Notes in
  Computer Science, pages 90–104. Springer, 2001}.

\bibitem{GMW}
O.~Goldreich, S.~Micali, and A.~Wigderson.
\newblock How to play any mental game or {A} completeness theorem for protocols
  with honest majority.
\newblock In {\em Proceedings of the 19th Annual Symposium on Theory of
  Computing ({STOC})}, pages 218--229, New York, NY USA, May 1987. ACM Press.

\bibitem{EigenTrust}
Sepandar~D. Kamvar, Mario~T. Schlosser, and Hector~G. Molina.
\newblock The eigentrust algorithm for reputation management in p2p networks.
\newblock In {\em Proceedings of the Twelfth International World Wide Web
  Conference, 2003.}

\bibitem{Agree80}
L.~Lamport, R.~Shostak, and M.~Pease.
\newblock Reaching agreement in the presence of faults.
\newblock {\em {J}ournal of the {ACM}}, 27(2):228--234, 1980.

\bibitem{Agree82}
L.~Lamport, R.~Shostak, and M.~Pease.
\newblock The byzantine generals problem.
\newblock {\em {ACM} Transactions on Programming Languages and Systems},
  4(3):382--301, 1982.

\bibitem{Fairplay}
D.~Malkhi, N.~Nisan, B.~Pinkas, and Y.~Sella.
\newblock Fairplay --- a secure two-party computation system.
\newblock In {\em Proc. Usenix Security Symposium 2004}, 2004.

\bibitem{Separable}
Damon Mosk-Aoyama and Devavrat Shah.
\newblock Computing separable functions via gossip.
\newblock In {\em PODC '06: Proceedings of the twenty-fifth annual ACM
  symposium on Principles of distributed computing}, pages 113--122, New York,
  NY, USA, 2006. ACM Press.

\bibitem{Paillier}
Pascal Paillier.
\newblock Public-key cryptosystems based on composite degree residuosity
  classes.
\newblock In {\em EUROCRYPT '99, Springer-Verlag (LNCS 1592)}, pages 223--238,
  1999.

\bibitem{BibDB:BookPearl}
J.~Pearl.
\newblock {\em Probabilistic Reasoning in Intelligent Systems: Networks of
  Plausible Inference}.
\newblock Morgan Kaufmann, San Francisco, 1988.

\bibitem{Pedersen91}
T.~P. Pedersen.
\newblock Non-interactive and information-theoretic secure verifiable secret
  sharing.
\newblock In {\em Proc. of CRYPTO 1991, the 11th Ann. Intl. Cryptology Conf.,
  Springer-Verlag (LNCS 576)}, pages 129--140, 1991.

\bibitem{SSS}
Adi Shamir.
\newblock "how to share a secret".
\newblock In {\em Communications of the ACM}, volume~22, pages 612--613, 1979.

\bibitem{DIMES}
Yuval Shavitt and Eran Shir.
\newblock Dimes: Let the internet measure itself.
\newblock {\em {ACM} SIGCOMM Computer Communications Review}, 35(5):71--74,
  2005.

\bibitem{FastAgree87}
S.~Toueg, K.~J. Perry, and T.~K. Srikanth.
\newblock Fast distributed agreement.
\newblock {\em {SIAM} Journal on Computing}, 16(3):445--457, June 1987.

\bibitem{PP3}
Li~Wan, Wee~K. Ng, Shuguo Han, and Vincent C.~S. Lee.
\newblock Privacy-preservation for gradient descent methods.
\newblock In {\em KDD '07: Proceedings of the 13th ACM SIGKDD international
  conference on Knowledge discovery and data mining}, pages 775--783, New York,
  NY, USA, 2007. ACM.

\bibitem{Yao}
A.~Yao.
\newblock Protocols for secure computations.
\newblock In {\em Proceedings of the 23rd Symposium on Foundations of Computer
  Science ({FOCS})}, pages 160--164. {IEEE} Computer Society Press, 1982.

\bibitem{PP2}
Sheng Zhang, James Ford, and Fillia Makedon.
\newblock A privacy-preserving collaborative filtering scheme with two-way
  communication.
\newblock In {\em EC '06: Proceedings of the 7th ACM conference on Electronic
  commerce}, pages 316--323, New York, NY, USA, 2006. ACM.

\end{thebibliography}
\end{document}